\documentclass{aa}

\usepackage{graphicx}
\usepackage{txfonts}

\begin{document}

\title{Plasma dynamics in the flaring loop observed by RHESSI.}

\titlerunning{Plasma dynamics observed by RHESSI.}
\authorrunning{T.~Mrozek et al.}

\author{T.~Mrozek\inst{\ref{inst1},\ref{inst2}}
\and R.~Falewicz\inst{\ref{inst2}}
\and S.~Ko{\l}oma\'nski\inst{\ref{inst2}}
\and M.~Litwicka\inst{\ref{inst1}}}

\institute{Space Research Centre, Polish Academy of Sciences, ul. Bartycka 18a, 00-716 Warszawa, Poland \label{inst1}
\and 
Astronomical Institute, University of Wroc{\l}aw, ul. Kopernika 11, 51-622 Wroc{\l}aw, Poland \label{inst2}}

\date{Received 1 January 2020/ Accepted 2 January 2020}

\abstract {Hard X-rays (HXRs) contain the most direct information about the non-thermal electron population in solar flares. {The approximation of the HXR emission mechanism (bremsstrahlung), known as the thick-target model, is well developed.} It allows one to diagnose the physical conditions within a flaring structure. The thick-target model predicts that in flare foot points, we should observe lowering of HXR sources' altitude with increasing energy.} {The foot point of HXR sources result from the direct interaction of non-thermal electron beams with plasma in the lower part of the solar atmosphere, where the density increases rapidly. Therefore, we can estimate the plasma density distribution along the non-thermal electron beam directly from the observations of the altitude-energy relation obtained for the HXR foot point sources. However, the relation is not only density-dependent. Its shape is also determined by the power-law distribution of non-thermal electrons. Additionally, during the impulsive phase, the plasma density and a degree of ionisation within foot points may change dramatically due to heating and chromospheric evaporation. For this reason, the interpretation of observed HXR foot point sources' altitudes is not straightforward and needs detailed numerical modelling of the electron precipitation process.} {We present the results of numerical modelling of one well-observed solar flare. We used HXR observations obtained by RHESSI. The numerical model was calculated using the hydrodynamic 1D model with an application of the Fokker-Planck formalism for non-thermal beam precipitation.} {{HXR data were used to trace chromospheric density changes during a non-thermal emission burst, in detail. We have found that the amount of mass that evaporated from the chromosphere is in the range of $2.7\times10^{13}-4.0\times10^{14}\rm{g}$. This is in good agreement with the ranges obtained from hydrodynamical modelling of a flaring loop ($2.3\times10^{13}-3.3\times10^{13}\rm{g}$), and from an analysis of observed emission measure in the loop top ($3.9\times10^{13}-5.3\times10^{13}\rm{g}$). Additionally, we used specific scaling laws which gave another estimation of the evaporated mass around $2\times10^{14}\rm{g}.$}} {{Consistency between the obtained values shows that HXR images may provide an important constraint for models - a mass of plasma that evaporated due to a non-thermal electron beam depositing energy in the chromosphere. High-energy, non-thermal sources' (above 20 keV in this case) positions fit the column density changes obtained from the hydrodynamical model perfectly. Density changes seem to be less affected by the electrons' spectral index. The obtained results significantly improve our understanding of non-thermal electron beam precipitation and allow us to refine the energy balance in solar flare foot points during the impulsive phase.}}


\maketitle 

\section{Introduction} \label{sec:intro}
The most straightforward information about a non-thermal electron population in solar flares {comes from} radio waves and hard X-rays (HXRs). During solar flares, {electrons} transport energy from the reconnection region to the chromosphere where they deposit it via Coulomb collisions with ambient plasma. This is the base of the well-known thick-target model \citep{brown1971}. {The emission mechanism and its relation to plasma density and an electron spectral index is well understood.}

One of the applications of the thick-target model is a  theoretical relation between the altitude of HXR foot point sources and energy of emitted radiation \citep{brown1975}. This relation results from the bremmstrahlung emission mechanism, and it depends on the column density and the non-thermal electron beam spectral index \citep{brown2002}. The relation was observed \citep{matsushita1992, aschwanden2002, mrozek2006} and used to obtain the density structure within a flaring loop \citep{fletcher1996, aschwanden2002}. A detailed analysis of HXR sources' locations allows one to diagnose the physical conditions within a flaring structure because non-thermal particles can be treated as a tool that probes plasma at various altitudes {\citep{kontar2010}}. {In this study, the HXR images were reconstructed with a 40~s integration time, thus, such a density vertical structure obtained from observations was averaged over time. It does not reveal the dynamics of plasma caused by chromospheric evaporation. The evaporating plasma velocities are typically observed in the range $100-700\;\rm{km/s}$ \citep{tomczak1997, nitta2012, sadykov2019}. This means that assuming a 40~s integration time, we average the plasma density changes along the path being from $4000\;\rm{km}$ to $30000\;{\rm km}$ long.}

Chromospheric evaporation is a natural consequence of heating the chromosphere by an electron beam. {It was first detected in soft X-ray (soft X-ray) spectra \citep{canfield1982, antonucci1982} as blue-shifted components in CaXIX and Fe XXV resonance lines. HXR observations may also reveal chromospheric evaporation. Namely, HXR sources have been observed to change their positions along a flare loop during the impulsive phase, which was interpreted as some evidence for chromospheric evaporation  \citep{milligan2006, liu2006}. The plasma moving upwards causes the rising of a column density which affects the height of thermalisation of non-thermal electrons. As a consequence, we expect that HXR emission sources occur at rising heights as the flare evolves. This interpretation is based on the assumption that the change of an HXR source position, for a given energy, is connected to plasma density only.} However, \cite{oflannagain2013}, assuming that the non-thermal part of a spectrum is as low as 6 keV, show that the main contributor to the movement of HXR sources is a spectral index variability. 

\cite{reep2016} made a series of magnetohydrodynamic (MHD) flare models, with non-thermal and thermal heating, and conclude that in pure thermal models an altitude distribution of HXR sources cannot be reproduced, and the non-thermal heating is favourable. {Assuming pure non-thermal heating, the analysis of the altitude of HXR sources may concentrate on the interplay between a density and a spectral index of a non-thermal electron beam. With a high time resolution, of the order of 10 s, we should be able to analyse both the time evolution of the density and the spectral index changes.} The problem is that we have to find a flare strong enough to enable us to reconstruct HXR images with very high energy and angular resolutions. The flare should be located close to the solar limb to minimise projection effects. Moreover, it should have as simple a morphology as possible, which is typically not the case for strong flares. 

In this paper, we present the results of an analysis of a flare that fulfilled the above criteria. It gave us a chance to investigate the positions of HXR sources in detail and compare them with hydrodynamic (HD) modelling results. Observational data are presented in Section~2, and the methodology is discussed in Section~3. The results are shown in Section~4, and the conclusions are given in Section~5.

\section{Data analysis}

\begin{figure}
\resizebox{\hsize}{!}{\includegraphics{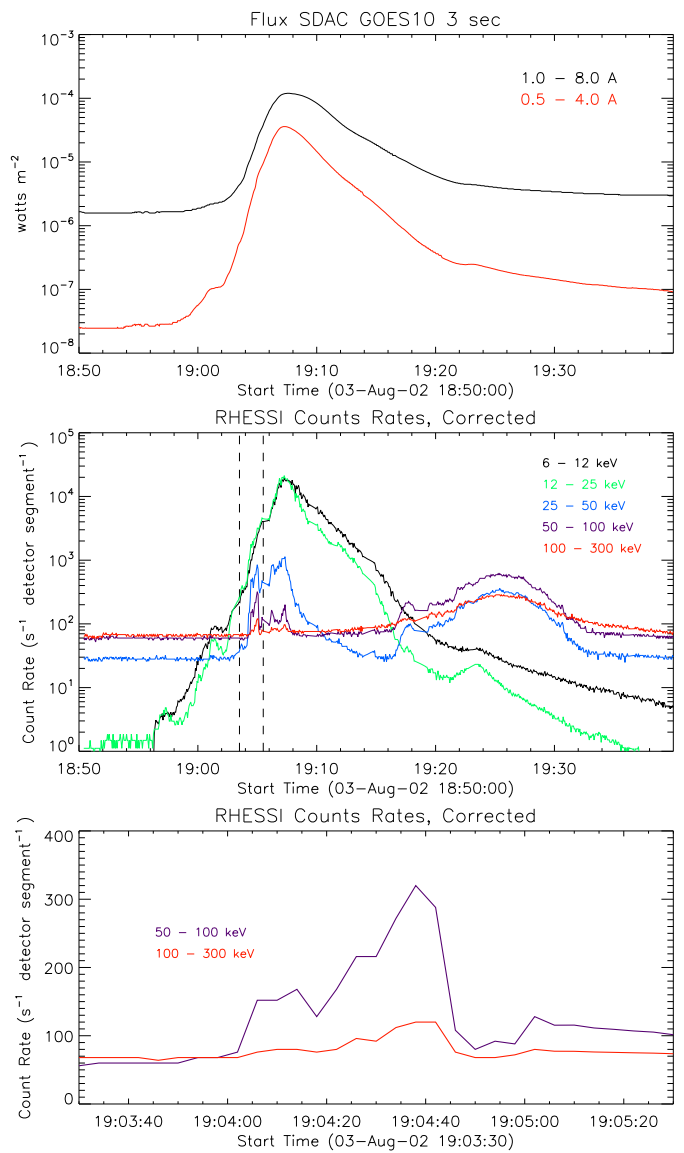}}
\caption{GOES (top panel) and RHESSI (middle panel) light curves for the SOL2002-08-03T19:07 flare. The bottom panel presents {the first strong} HXR burst observed from 19:04~UT to 19:05~UT {on which our analysis focuses} (the interval is marked in the middle panel with dashed vertical lines).}
\label{fig:GOES_RHESSI}
\end{figure}

The analysed flare, SOL2002-08-03T19:07, occurred close to the west limb of the solar disk (S15W70). It was a strong flare, of the GOES class X1.5 (Figure~\ref{fig:GOES_RHESSI}), with a complex morphology. {We used data obtained by four instruments. The Ramaty High Energy Solar Spectroscopic Imager \citep[RHESSI,][]{lin2002} was an imaging spectrometer equipped with nine germanium detectors. The detectors were able to register photons of a wide energy range from 3 keV up to 20 MeV, with an excellent 1 keV resolution below 100 keV \citep{smith2002}. This huge range was achieved via the utilisation of several systems, which reduced count rates in low energies. Among them were attenuators whose transmission is well understood at present, thus measurements made in various attenuator states may be analysed and compared quantitatively, which is important for our analysis. The Transition Region and Coronal Explorer \citep[TRACE,][]{handy1999} was a telescope equipped with extreme ultraviolet (EUV) and ultraviolet (UV) filters. The telescope had a great angular resolution, close to 1 arc sec. In our analysis, we used TRACE 171~{\AA} images as context data. The Extreme ultraviolet Imaging Telescope \citep[EIT,][]{delaboudiniere1995} was an ultraviolet telescope installed on board the Solar and Heliospheric Observatory \citep[SOHO,][]{domingo1995}. The images obtained by EIT were used for TRACE pointing correction. The GOES X-ray Sensor (XRS) is a monitor of solar X-ray flux which measures the total flux in two wavelength ranges. We use GOES light curves as context data.
}

RHESSI registered the entire event (Figure~\ref{fig:GOES_RHESSI}, middle panel). A bump seen in light curves during the decay phase (around 19:25 UT) is related to the radiation belts' passage. It does not influence the analysis since we were concentrated on the impulsive phase only when the satellite was outside the radiation belts. The impulsive phase occurred at 19:04-19:08~UT when strong HXR bursts are clearly visible in higher energies. 

The imaging capability of RHESSI was achieved with nine pairs of grids (periodic system of slits and slats) and through rotation of the entire satellite \citep{hurford2002}. This produced modulation of incoming HXR flux. Knowing the geometry of the system (slits and slats sizes, and distances), we could reconstruct an actual distribution of sources with several algorithms available in the Solar Soft Ware (SSW) library. {RHESSI grids have a spatial resolution from $2.26$~arc sec (grid No.~1) to $183.2$~arc sec (grid No.~9). The image spatial resolution depends on the grids chosen for reconstruction. In practice, RHESSI image reconstruction is performed for grid Nos. 3–6, 8, and 9. Depending on the reconstruction algorithm and weights chosen, this set of grids gives a spatial resolution of about $7-9$~arc sec \citep{aschwanden2002}. The spatial resolution strongly depends on the sources' distribution and count rates, so sometimes, grid No.\ 1 may be used for image reconstruction giving the intended spatial resolution well below 7~arc sec \citep{liu2007}}

In our work, we used images reconstructed using the CLEAN \citep{hogbom1974} and PIXON \citep{pina1993} algorithms for grid Nos. 3–6, 8, and 9 with uniform weighting. CLEAN gives precise locations of HXR sources, while PIXON is very well-suited for photometry \citep{aschwanden2004, chen2012}. However, PIXON is very slow compared to CLEAN; therefore, we used CLEAN to produce hundreds of images for a quick look. The optimal imaging parameters (time and energy ranges) estimated from CLEAN images were used to reconstruct images with the PIXON algorithm. We checked the locations of sources visible in the images reconstructed with both methods. We estimated the mean difference between centroids of CLEAN and PIXON sources to be 1.1~arc~sec. Therefore, we utilised PIXON images only as they gave more compact, not over-resolved sources which can be easily spatially separated for imaging spectroscopy.

{The impulsive phase of the analysed flare was very strong, which allowed us to reconstruct RHESSI images with high time resolution and narrow energy intervals. For the overall analysis, we reconstructed images for twelve time intervals covering 19:03--19:06~UT. The first two intervals, before the main peak, were important for the proper start of the modelling procedure. Next, five 12-s long intervals were used for a detailed analysis of the energy-altitude relation. The remaining five intervals (until 19:06~UT) were used to conduct modelling to the flare maximum. However, it has to be stressed that after 19:05:12~UT, RHESSI's attenuator state changed (the second one was inserted) which added more uncertainty to low-energy counts. For this reason, we do not discuss images and spectra obtained after 19:05:12~UT quantitatively. Nevertheless, in time plots, we present all data points obtained up to 19:06~UT.} We defined narrow energy bands with widths increasing with energy. Up to 20~keV we were able to use $ \Delta E = 2~keV $, while above 100~keV the widths were  $ \Delta E = 20~keV $, which is still high energy resolution for imaging with the 12~s time resolution. The shift between neighbouring energy ranges was equal to half of their width. This gave us more data points for the spectral fitting with OSPEX.

\begin{figure}
\resizebox{\hsize}{!}{\includegraphics{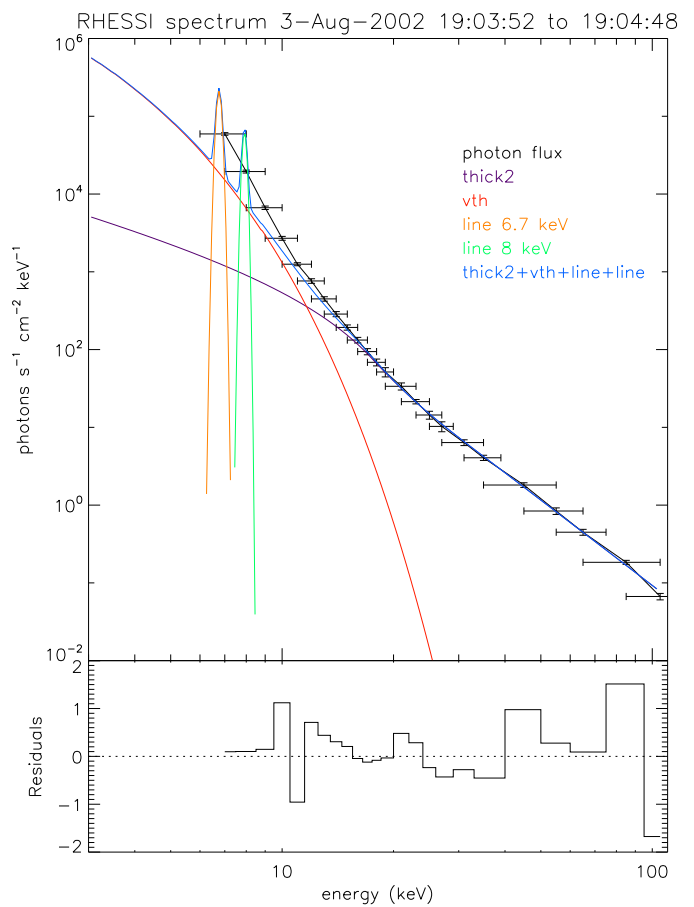}}
\caption{{Example of fit to RHESSI spectrum. Top panel: RHESSI photon flux (error bars) registered from the area marked in Figure~\ref{fig:TRACE_RHESSI_hi_res} during the entire 19:04 -- 19:05~UT peak. Components of the fit used are as follows: thermal continuum (red), thick-target model (violet), and the Fe (6.7~keV) and Ni (8.0~keV) line complexes (orange and green Gaussians, respectively). The sum of the components is presented with the light blue line. Bottom panel: Normalised residuals (sigma) for the spectral fit.}}
\label{fig:spectrum_fit_example}
\end{figure}

\begin{figure}
\resizebox{\hsize}{!}{\includegraphics{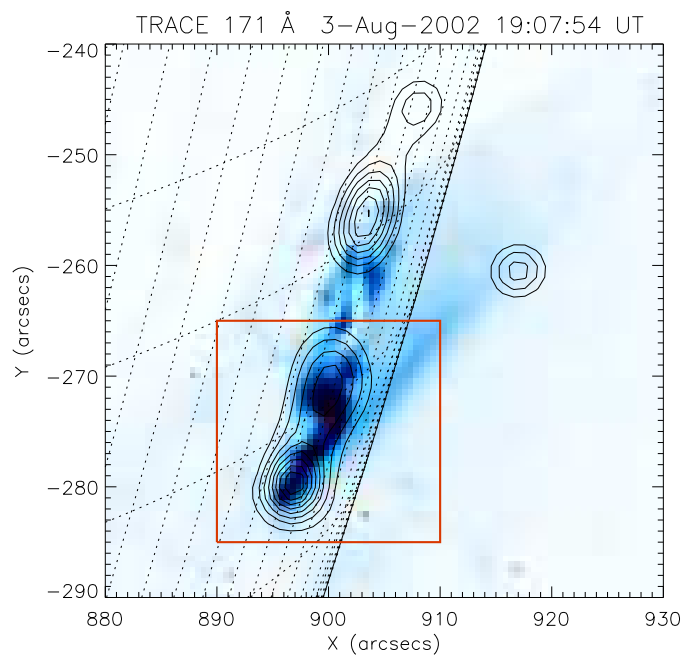}}
\caption{TRACE 171~\AA\mbox{} image with overlaid contours of reconstructed RHESSI sources (19:04:28-19:04:40~UT, 31-36~keV, PIXON). The area covering the modelled loop is marked with the red square.}
\label{fig:TRACE_RHESSI_hi_res}
\end{figure}

RHESSI spectra were analysed with the OSPEX package,  which is available in IDL's SSW library. Figure~\ref{fig:spectrum_fit_example} shows an example of the spectrum fit for a HXR signal measured in the area marked in Figure~\ref{fig:TRACE_RHESSI_hi_res} {integrated over the entire strong peak} (19:04 -- 19:05~UT). {We performed fits using three components. Firstly, a  thermal component was used to fit the thermal continuum usually observed up to several kiloelectron volt. We assumed that the observed thermal emission comes from isothermal plasma. Secondly, Gaussians were used to fit the line complexes present around 6.7~keV and 8.0~keV. This emission comes from highly ionised Fe and Ni. Thirdly, a non-thermal component was fitted with the bremsstrahlung thick-target model. It gave us a set of parameters describing the non-thermal electron beam which were further used as input parameters for model calculations.

}

We used the same set of functions for every spectrum analysed in this paper. The example shown in Figure~\ref{fig:TRACE_RHESSI_hi_res} was made for 60~s time integration, thus we had good statistics and the spectral fit easily converges to low $\chi^2$ and almost random residuals. However, we have to remember that for lower count statistics cases (12-s long time integration) the fit is very sensitive to start values of parameters and the interval in which we allow them to change. It was especially important for low energies where we used thermal continuum and two Gaussians. {Therefore, we kept the iron abundance relative to hydrogen fixed and allowed the centroid of Gaussians to change only by $\pm 0.3$~keV. }In the non-thermal function, we kept the energy break (1500~keV) and high energy cut-off (32000~keV) fixed. {Other parameters of non-thermal fit, which we allowed to change in the nominal OSPEX ranges, are the following: the total integrated electron flux  $[10^{35}\;\rm{\frac{electrons}{s^{-1}}}]$,\;$10^{-10}-10^{10}$; the power-law of the electron flux distribution function below break energy, $1.1-20$; the power-law index of the electron flux distribution function above break energy,  $1.1-20$; and the low-energy cut-off in the electron flux distribution function, $1-1000$~keV.

}

RHESSI images, reconstructed for the strongest HXR burst, revealed up to four {HXR foot point sources (Figure~\ref{fig:TRACE_RHESSI_hi_res})}. {The HXR images revealed the complexity of the event, which was further investigated with EUV data using the TRACE observations.} {TRACE images were corrected for pointing inaccuracies using a standard method, namely the correlation with SOHO/EIT images\footnote{\url{https://www.tcd.ie/Physics/people/Peter.Gallagher/trace-align/index.html}}.} In the case of SOL2002-08-03T19:07, images made with the same filters were not available. Therefore, we compared TRACE 171~{\AA} to EIT 195~{\AA} images, thus the accuracy of the pointing correction may be worse. To obtain as good a pointing correction for TRACE as possible, we calculated offsets between EIT and TRACE several times, covering various phases of the flare. In each case, we obtained a similar shift; therefore, we assumed that spectral differences between the two filters {do not affect} the accuracy of the method. The overall spatial correlation of TRACE and RHESSI images is not better than 5~arc sec. Nevertheless, it is enough to recognise individual structures, which helped us to understand the flare morphology.

\begin{figure}
\resizebox{\hsize}{!}{\includegraphics{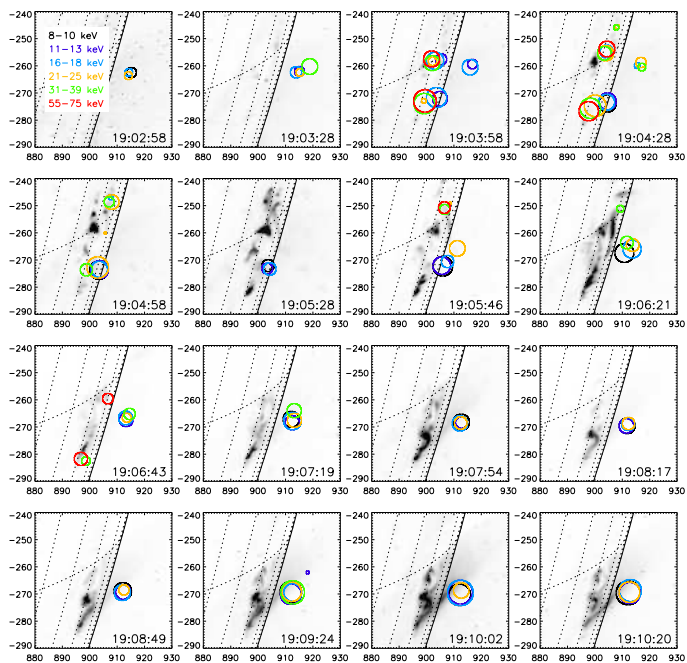}}
\caption{{TRACE 171~\AA\mbox{} images with over plotted contours (70\% of maximum brightness) of reconstructed RHESSI sources (PIXON) for a few energy ranges. Energy is colour-coded (top-left panel inset). See the text for further details.}}
\label{fig:TRACE_RHESSI}
\end{figure}

Figure~\ref{fig:TRACE_RHESSI} presents TRACE 171~{\AA} images covering the {19:02:58--19:10:20~UT} time interval. The EUV images revealed two main structures that may be recognised. The first was a small and compact loop observed around 19:08~UT. The second was a larger arcade visible in a later phase.

The overall morphology was too complicated to be approximated as a single loop model, which was our aim. Thus, we decided to concentrate on the first strong HXR burst {(see Figure~\ref{fig:GOES_RHESSI}, the bottom panel}), connected with the small loop clearly seen around 19:08~UT. The loop decayed very quickly and was barely visible three minutes later (19:11~UT). In the TRACE images, we overlaid {intensity isolines} of RHESSI sources reconstructed in relevant times {(Figure~\ref{fig:TRACE_RHESSI})}. The sources were reconstructed in several energy bands covering the range of 6-120~keV, but, for clarity, we present only a few energy intervals {ranging from 6-10~keV to 55-75~keV.}

It has to be stressed that, in general, we have to deal with two different types of sources when comparing EUV and HXR images. {bf Firstly, non-thermal sources which are emission from foot points, in most cases. HXR radiation and EUV radiation of a foot point occur almost simultaneously. Secondly, thermal sources located in the top of flaring loops. These structures that are visible in HXR have to cool down to be visible in EUV. It usually takes hundreds of seconds, so a structure seen in EUV at a given time should be compared with HXR sources that are visible earlier. 
}

{Figure~\ref{fig:TRACE_RHESSI_hi_res} presents the EUV image taken at 19:07:54~UT with overlaid contours of RHESSI sources reconstructed during the main HXR burst. In that case, we used the PIXON algorithm, chose the energy range  31-36~keV, and used the integration time of 12~s {(19:04:28-19:04:40~UT)}. High count rates allowed us to reconstruct RHESSI images with an angular resolution approximately equal to 5 arc sec. Two foot points related to the small loop are clearly visible. Other visible HXR sources were connected to the large arcade, so they were not taken into consideration for the single loop hydrodynamical modelling.}

\section{Energy-altitude relation} {\label{sec:3}}

We expect that foot point sources will decrease their altitude in the corona with the rising energy of non-thermal electrons. This is a direct consequence of the bremsstrahlung emission mechanism. In general, the observed energy-latitude relation depends on an actual power-law index of an electron spectrum, a density distribution within a flaring loop, and an ionisation state of plasma. All these parameters may change during the impulsive phase, and the changes may be complicated. 

{\subsection{Construction of the reference level}}

\begin{figure}
\resizebox{\hsize}{!}{\includegraphics{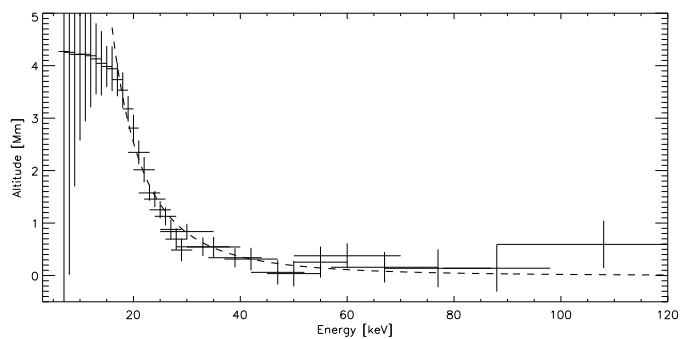}}
\caption{Energy-altitude relation for the HXR sources connected with the small loop marked with the red square in Figure~\ref{fig:TRACE_RHESSI_hi_res}. The sources were reconstructed for the entire analysed HXR burst (19:04 -- 19:05~UT). {The power-law fit is presented with the dashed line. According to Equation~\ref{eq:E-A_fit}, the fitted parameters are as follows: $z_0 = 2.53\,{\rm Mm}$ and $a=0.36$.}}
\label{fig:E-A_main_peak}
\end{figure}

{To construct the reference level, we used images that were reconstructed for the entire 60~s long HXR peak. It allowed us to avoid a problem with low count statistics for high energy sources. The reference level was determined by such sources, thus, we had to reconstruct them with high precision. As the first approximation, we defined a provisional reference level as a line connecting centroids of the foot point sources visible above 50~keV. Next, for each source, we measured its distance from the provisional reference level, which defined the altitude of the source. 

The altitudes of the sources related to the small loop are presented in Figure~\ref{fig:E-A_main_peak}. The errors of the centroids' locations were estimated according to \cite{bogachev2005}. {Large errors seen for low energies result from a large and diffuse source visible below several kiloelectron volts.} This part of the spectrum is purely thermal, so we excluded it from the altitude analysis. Above 15~keV, we can clearly see decreasing of the sources' altitudes with rising energy. This is consistent with the previous observations of the energy-altitude relation \citep{matsushita1992, aschwanden2002}. 

According to \cite{aschwanden2002}, the energy-altitude relation was fitted with the following power-law function:
\begin{equation}
    z\left(\epsilon\right)=z_0\left(\frac{\epsilon}{20\,\rm{keV}}\right)^{-a}
\label{eq:E-A_fit}
,\end{equation}
where $z$ is an altitude, $a$ is a power-law index, and $z_0$ is the altitude of a 20~keV source. The fit to purely non-thermal sources is presented in Figure~\ref{fig:E-A_main_peak}. {The fitted parameters are as follows: $z_0 = 2.53\,{\rm Mm}$ and $a=0.36$.} Assuming a zero pitch angle and neglecting pitch angle scattering, we can define a column density $N$ needed to stop an electron of energy $E$ \citep{brown2002}:
\begin{equation} 
    N\left(E\right)=\frac{E^2}{2K}
\label{eq:col_den}    
,\end{equation}
where $K$ is a constant \citep{spitzer1962}. In the simplest case, we may assume that electrons of energy $E$ produce (via collisions) photons of energy $\epsilon\approx E$. Therefore, we may write the following:
\begin{equation}
     N\left(\epsilon(z)\right)=\frac{\epsilon^2}{2K}
 \label{eq:col_den_epsilon}      
.\end{equation}
Derivating $N$ with regard to $z$, we get the relation between the number density ($n$) and the altitude:
\begin{equation}
    n\left(z\right) = -\frac{dN(z)}{dz} = -\frac{\epsilon}{K}\frac{d\epsilon}{dz}    
.\end{equation}
From Equation~\ref{eq:E-A_fit}, we can calculate $\epsilon$ and its derivative $\frac{d\epsilon}{dz}$. Inserting these values into Equation~\ref{eq:col_den}, we get the following:
\begin{equation}
    n\left(z\right)=n_0\left(\frac{z}{z_0}\right)^{-1-\frac{2}{a}}
\label{eq:density_distribution}
,\end{equation}
where:
\begin{equation*}
    n_0 \approx 1.5\times10^{12}\left(\frac{1}{a}\right)\left(\frac{1\,\rm{Mm}}{z_0}\right)\rm{cm^{-3.}}
\end{equation*} 
The altitude of the photospheric density ($1.16\times10^{17} \rm{cm^{-3}}$) was found from such a density-altitude relation and used as the final reference level allowing us to compare the observed altitudes with the modelled ones. The reference level we used ({$0.46\,\rm{Mm}$}) is constant in time.} 

{\subsection{Time evolution of energy-altitude relation}}

Figure~\ref{fig:E-A_main_peak} was obtained for a strong, long-lasting (almost 60~s) HXR burst, meaning that this figure shows an averaged relation. We can expect the energy-altitude relation changes seen in Figure~\ref{fig:E-A_main_peak} due to a changing electron beam spectrum and plasma density evolution during these 60~s. {From our set of images reconstructed in short-time intervals (12~s long), we chose five consecutive intervals covering the strong peak, which enabled us to investigate the time evolution of the energy-altitude relation. }

The set of images obtained in different energy ranges and different time intervals was used for further analysis. For each foot point, we determined the location of its centroid and measured a distance from the reference level (altitude). {Moreover, at this stage we checked the difference between CLEAN and PIXON altitudes again. The mean value was $-0.23\,\pm\,0.7$~Mm.}

The analysed loop is small, and its foot points do not separate on every RHESSI image. We reconstructed more than 1000 images with changing reconstruction parameters. {These parameters were weights and grids used for image reconstruction. Both of them influence the final spatial resolution of images. Namely, using grid No. 3 as the finest, we could slightly change the final resolution by changing a weighting factor to signal modulation from this grid. The grids used for RHESSI image reconstruction also affect the final image's resolution. If we use the finest grid whose resolution is finer than the real size of the observed source, then we add only noise to the final image \citep{kolomanski2011}. Therefore, to get the best final set of images, we had to use both weighting schemes (natural or uniform) and test grids Nos. 3 and 4 as the finest. As we used ten time intervals in total and 30 energy ranges, the number of the reconstructed PIXON images easily outnumbered 1000 images.} In each image revealing two foot points, we searched for any asymmetries in foot points' characteristics. We concluded that the altitude and brightness of both foot points are very similar; there is almost no asymmetry. Therefore, we decided to analyse the energy-altitude relation without separating foot points. This enabled us to include images where sources were visible and not separated. 

\begin{figure}
\resizebox{\hsize}{!}{\includegraphics{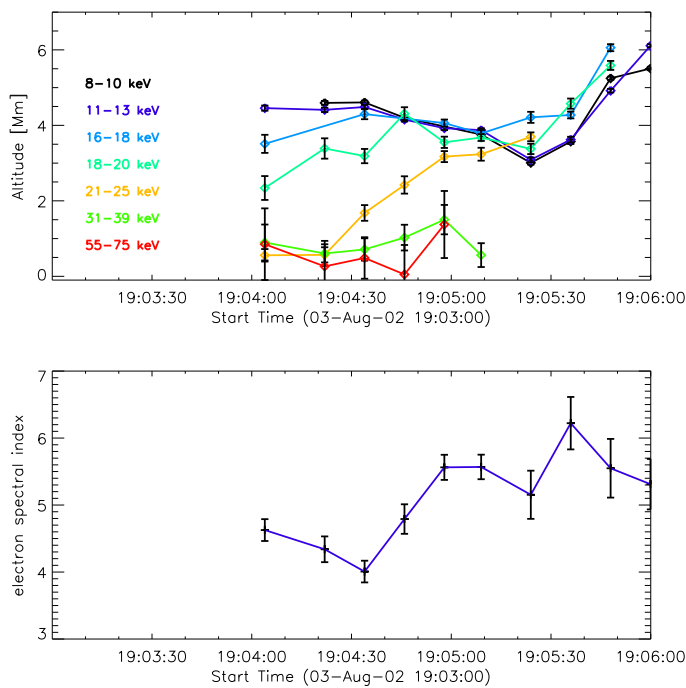}}
\caption{{Source altitude changes with time.} Top panel: Sources' altitudes change in time for several energies (colour-coded). Bottom panel: Electron spectral index changes derived from spectra registered in relevant sources.}
\label{fig:E-A_delta}
\end{figure}

Figure~\ref{fig:E-A_delta} (top panel) presents an example of the sources' altitude changes in time and for several energies. In low energies, we observed the source close to the loop top with the stable altitude. Among the presented curves, the most variable is the 21-25~keV one. At the beginning of the HXR burst, this source is located close to the reference level, while 30 seconds later the emission in this range is observed close to the loop top. The spectra analysis revealed that the non-thermal component dominates above 13-15~keV, thus, we may conclude that in 21-25~keV, we have non-thermal emission only. {Altitudes change less for higher energies (31-39~keV and higher).  We observed a small increase in altitudes around 19:05~UT, but this is within measured uncertainties.} 

The observed evolution {of HXR sources' positions} shows a lot of dynamic changes in the non-thermal electron precipitation region if we assume that these changes are density-related only. However, it is known from theoretical calculations that the index of the power-law spectrum of non-thermal electrons may also influence the observed altitudes of non-thermal emission sources. We were able to extract spectral information for individual sources using imaging spectroscopy. The bottom panel of Figure~\ref{fig:E-A_delta} presents non-thermal electron spectral indexes for the same time intervals as presented in the top panel. We can qualitatively compare power-law index changes with the evolution of HXR sources' positions. It is seen that, for some energy ranges {(18-20~keV, 21-25~keV)}, an increase in altitude started while the spectrum was still hardening. This suggests that density changes may dominate in this case. To verify this hypothesis, we performed numerical hydrodynamical modelling of the loop with electron beams as an energy transport mechanism.

\section{Modelling the small loop}

We assumed the electron beam-driven evaporation model of a solar flare. The distribution of plasma in the small loop was modelled with the use of the modified Naval Research  Laboratory solar flux tube model \citep{mariska1982}. The separate module of our code was used to calculate electron distribution in the loop as a function of a position along a magnetic field line, an electron pitch angle, and electron energy. For this purpose, we used the Fokker-Planck formalism (McTiernan and Petrosian 1990) and the open-source code prepared by G.~Holman\footnote{ \url{hesperia.gsfc.nasa.gov/hessi/flarecode/efluxprog.zip}}. {A detailed diagram of the procedure and the assumptions is described in \cite{falewicz2014}.}  

{The loop's parameters (height, foot point separation) were estimated based on RHESSI images. We assumed a semi-circular loop whose half-length was determined by the separation of the centroids of its foot points. For this purpose, we reconstructed a RHESSI image in the 35-55~keV energy range and the 19:04 -- 19:05~UT time range with the PIXON algorithm. Thus, we obtained the image with very good count statistics and well-resolved sources, which was used for a loop parameters estimation. We assumed that uncertainty of the measured sources' positions, of the order of 1 arc sec, is similar to the difference between the positions measured in CLEAN and PIXON images. A half-length of the loop, $\rm{L}_0$, {was estimated from the distances between the centroids of the foot points in the PIXON image}, assuming a semi-circular loop shape. The $\rm{L}_0$ was set to $6.7\times10^8\,\rm{cm}$. {The loop cross-section, S, was calculated from the foot-point plane-of-image area. First, we determined the area within a $50\%$ intensity isoline for each foot point separately. Then we calculated the average value of the two areas because we did not observe a substantial foot points' asymmetry.} Finally, S was set to  $5.2\times10^{16}\,\rm{cm}^2$.} A gas pressure value (P) at the base of the transition region for the beginning of the flare modelling was assumed to be $\rm{P_0} = 32\,\rm{dyn\,cm^{-2}}$. This value was estimated from the equalisation of the modelled and observed $6-10$~keV fluxes for the beginning of the flare (the first time interval) with the methodology described below.

{The model's thermal emission was defined by a single temperature and emission measure of the optically thin thermal plasma, and it is based on the X-ray continuum and line emission calculated by the CHIANTI atomic code \citep{dere1997, landi2006}. For plasma temperatures above $10^5$~K, the coronal element abundances were used \citep{feldman2000}, while below $10^5$~K, photospheric abundances were applied. The thick-target emission was defined by the total integrated non-thermal electron flux, $\rm{F_{nth}}$, the power-law index of the electron energy distribution, $\rm{\delta}$, and the low-energy cut-off of the electron distribution, $\rm{E_c}$. These parameters were then used as characteristics of the electrons injected at the top of the modelled loop. The energy deposition rates were calculated using an approximation given by \cite{fischer1989}. 

{The loop was modelled from the beginning of the flare (19:03~UT) to the $6-10$~keV X-ray brightness maximum (19:06~UT). Steady-state spatial and spectral distributions of the non-thermal electron beam along the flaring loop were calculated for each time step of the model using the Fokker–Planck formalism. Using these data, spatial distributions of the thermodynamic parameters of the flaring plasma, X-ray thermal and non-thermal emissions, and the integral fluxes in the selected energy ranges were calculated for each time step. {The time step of the non-thermal electron beam was defined as the accumulation time needed to obtain the spectrum from RHESSI data with a sufficient signal-to-noise ratio (S/N). For this reason, during each non-thermal electron beam’s time step, heating of the loop was calculated with fixed non-thermal electron beam parameters, while the thermodynamic parameters of the flaring plasma varied, being calculated with a much shorter time step of the HD model.} 

The modelling procedure for each time interval was performed as follows:
\begin{enumerate}
    \item We fitted the RHESSI spectrum with a set of analytical functions (thermal, two Gaussians, and non-thermal).
    \item We estimated the total observed flux in the $6-10$~keV interval. 
    \item We used fitted non-thermal electron spectrum parameters to calculate the energy input via the Fokker-Planck equation. \label{energy_input}
    \item We calculated the flux in the $6-10$~keV interval from the model and compared it with the observed one.
    \item If the difference wass larger than 1\%, then we changed the cut-off energy, $\rm{E_{c}}$, and returned to the step \ref{energy_input} \label{ecutoff_optimization}
    \item If the difference between the observed and the modelled fluxes was below 1\%, then we moved to the next time interval.
\end{enumerate}

\begin{figure}
\resizebox{\hsize}{!}{\includegraphics{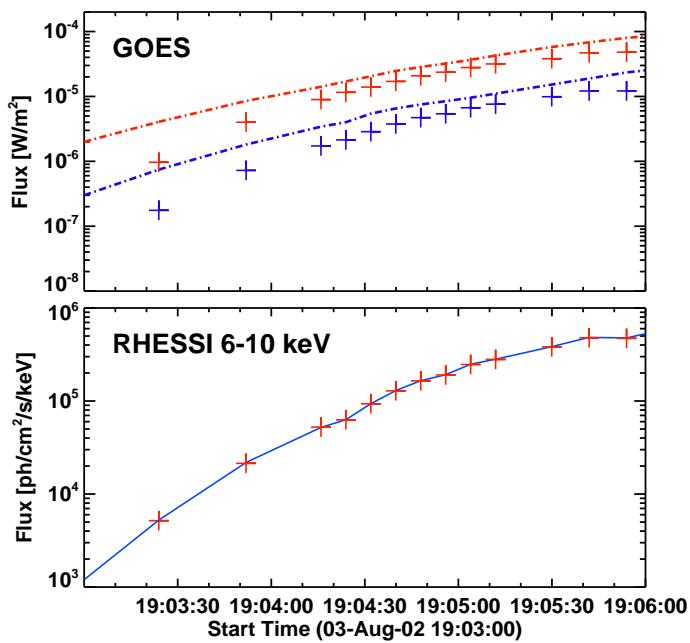}}
\caption{{ Modelled X-ray fluxes (plus signs) compared to GOES (top panel), and to RHESSI 6-10~keV (bottom panel). Obtained values are slightly lower than the GOES flux as we modelled only the emission coming from the fraction of the solar surface. The agreement between the model and RHESSI fluxes is the  method constraint - we were changing the cut-off energy until agreement was achieved.}}
\label{fig:model_goes_rhessi}
\end{figure}

\begin{figure}
\resizebox{\hsize}{!}{\includegraphics{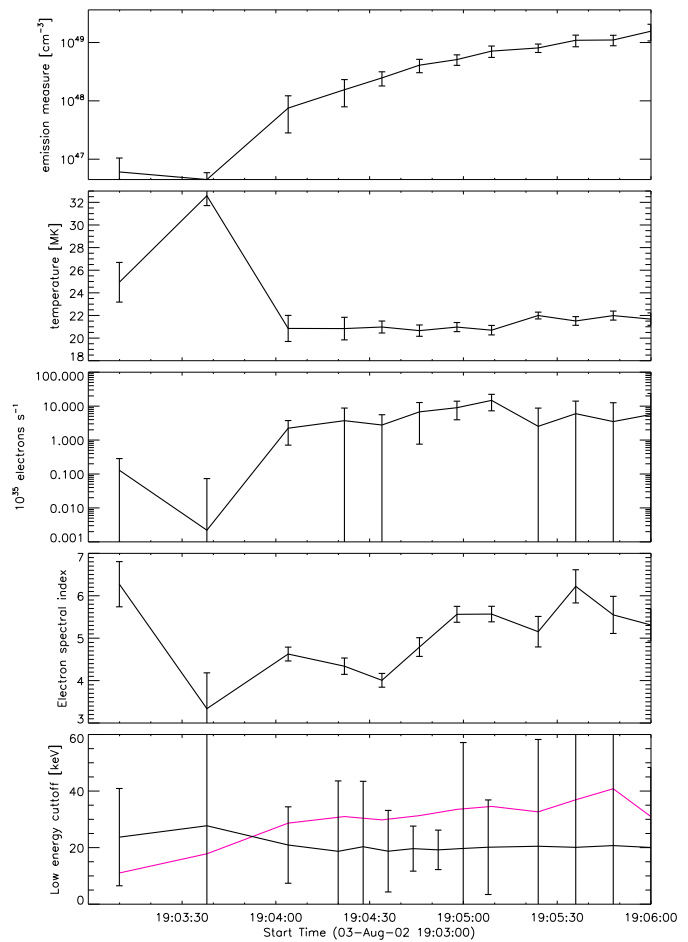}}
\caption{{ Spectral fit results (parameters marked with black error bars) for the time interval 19:03-19:06~UT. In the last panel, $E_c$ obtained from the spectral fit is compared to $E_c$, which was adjusted in the modelling procedure (pink line).}}
\label{fig:ospex_params}
\end{figure}

The iteration described in steps \ref{energy_input}--\ref{ecutoff_optimization} was done because the energy input via non-thermal electrons is very sensitive to  $\rm{E_c}$. A variation in this value of just a few {\bf kiloelectron volts} can add or remove a substantial amount of energy to or from a flare loop because of the power-law nature of the energy distribution. Thus, $\rm{E_c}$ must be selected with a high accuracy to achieve the agreement between a model and observations. In our previous works \citep{falewicz2009, falewicz2011, falewicz2014, siarkowski2009}, we used GOES data as a reference X-ray flux for model calculations. In this work, we analysed only a part of the entire X-ray emission coming from the small loop. Other X-ray sources existed on the solar disk at the same time; therefore, we could not use GOES which integrates signals from the whole visible solar disk. Therefore, we fitted spectra to the flux summed over the sources visible within the area marked with the red box in Figure~\ref{fig:TRACE_RHESSI_hi_res}. {Modelled fluxes compared to GOES and RHESSI are presented in Figure~\ref{fig:model_goes_rhessi}.} {The obtained fit parameters are presented in Figure~\ref{fig:ospex_params}}. After a spectral fitting, we calculated the observed integral photon flux in the range $6-10$~keV. The $\rm{E_c}$ in the model was carefully adjusted until an agreement with the observed flux was achieved. {The bottom panel of Figure~\ref{fig:ospex_params} presents the adjusted $\rm{E_c}$ (pink line). Usually, the optimised value of $\rm{E_c}$ was bigger than the fitted one of about $2-20$~keV. However, for almost each data point, the adjusted $\rm{E_c}$ was within the uncertainties of $\rm{E_c}$ from the spectral fitting.} Next, the procedure was applied for consecutive time steps. }}

{\section{Discussion}}

\begin{figure}
\resizebox{\hsize}{!}{\includegraphics{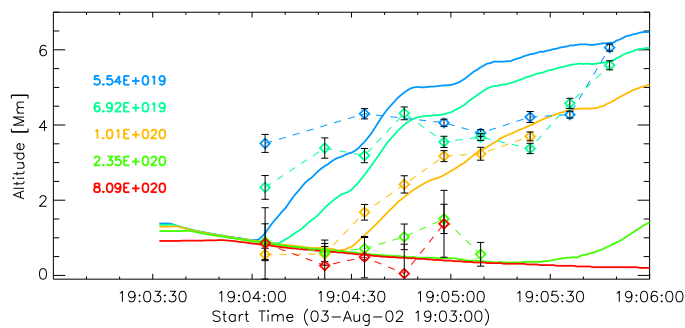}}
\caption{{Solid lines present column density evolution as derived from model calculations. The lines are presented for several values ($\rm{cm^{-2}}$) listed in the legend. {Additionally we plotted locations of HXR sources (diamonds and dashed lines). The colour-coding relates energies with relevant column densities.  Colours are the same as in Figure~\ref{fig:E-A_delta}, the red line is for $55-75\,\rm{keV}$, yellow presents $21-25\,\rm{keV}$, etc. }}}
\label{fig:E-A_col_den}
\end{figure}

{Figure~\ref{fig:E-A_col_den} shows the column density evolution derived from the hydrodynamical modelling of the observed loop. We plotted five curves that show the time evolution of altitudes for certain constant levels of the column density. {Utilising Equation~\ref{eq:col_den_epsilon}, we can relate a column density with the energy of a non-thermal electron that is stopped for a given value of the column density. Therefore the altitudes }of the HXR sources from Figure~\ref{fig:E-A_delta} can be over-plotted since we relate source energy with column density. In such a way, we compare the modelled column density with the observed one. There is a very good agreement between the column density and the altitudes of purely non-thermal emission, which supports the scenario that the observed altitudes are mainly related to density changes. However, for the energy intervals 16-18~keV and 18-20~keV, and for the first two time intervals, the observed column densities are above the modelled ones. The effect might be related to the depth of non-thermal electron energy deposition. \cite{mrozek2007} analysed the relation between the power-law index and thermal emission productivity for SXR, EUV, and UV foot point sources. The productivity was defined as a ratio of the SXR (or EUV or UV) signal to the HXR flux measured for a given foot point. The authors found that the SXRs' productivity rises with a rising power-law index (softer spectrum). For EUVs, the relation was almost flat, while for UVs there was a negative correlation (more UVs produced by lower values of a power-law index). Taking this result into account, we can explain the difference between the modelled and observed column densities seen for intervals 16-18~keV and 18-20~keV. For the first two analysed time intervals, the non-thermal spectrum was softer than for the next one, which means that relatively more energy is contained in low-energy electrons that deposit their energy mainly in the SXR emitting region. Therefore, we may expect that some additional thermal emission is present close to (above) 16-20~keV sources, which may slightly change (rise) the sources' altitudes. We may expect that a power-law index is mainly responsible for observed altitude changes of low-energy sources. This is supported by \cite{oflannagain2013} who analysed altitude changes of very low-energy (below 10~keV) non-thermal sources. They conclude that the observed changes of HXR sources' altitudes may only be explained by changes of a power-law index. We may see a trace of such a scenario at the beginning of the analysed HXR pulse. However, when a  non-thermal spectrum is hard enough, and a large fraction of electrons deposit their energy deeper in the chromosphere, then the dominant in affecting HXR sources' altitudes would be density changes.} This supports earlier observations of HXR moving sources in RHESSI images, which was interpreted as chromospheric evaporation \citep{milligan2006, liu2006}. For the source reconstructed in the 21-25~keV energy range, we estimated the evaporation velocity to be almost 150~km/s. 

\begin{figure}
\resizebox{\hsize}{!}{\includegraphics{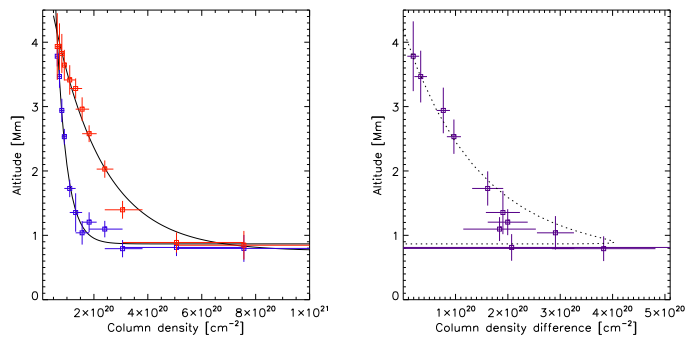}}
\caption{{{Column density changes.} Left panel: Column density - altitude relations for the start (blue) and the end of analysed burst (red). The relations were smoothed with over a three-point window. The exponential fits are presented with black lines. Right panel: Column density difference obtained by subtraction of points (red - blue) from the left panel (violet points). The difference between exponential fits is presented with a dotted line.}}
\label{fig:column_density_changes}
\end{figure}

Assuming that purely non-thermal HXR sources precisely indicate a column density value at some level, we may construct a method to estimate the total mass evaporated during the impulsive phase directly from HXR observations. Subtracting the two energy-altitude relations obtained at the beginning and at the end of a HXR burst, we get a curve describing the density change in the non-thermal electron precipitation region. {As described by Equations~\ref{eq:E-A_fit} and \ref{eq:col_den_epsilon} in Section~\ref{sec:3}, under certain assumptions, a relation between the column density and the altitude can be used.} An example for the main burst of SOL2002-08-03T19:07 is presented in Figure~\ref{fig:column_density_changes} (left panel). The column density - altitude relations for the start and the end of the analysed burst {(19:04-19:05~UT)} are shown {with the blue and red points}, respectively.
 
Subtraction of both curves informs us about additional mass that occurs above a given altitude. The maximum of a column density change (the most right point on the curve) allowed us to estimate the total mass that was pushed upward (above the given altitude). In the analysed case, we determined that above the altitude of 1000~km (above the photosphere), the change of the column mass is almost $4\times10^{20}\,\rm{cm^{-2}}$. Assuming a constant loop cross-section, we estimated the mass pushed upward to be  $7.3\times10^{13}\,{\rm g}$. 

The energy-altitude relation may be noisy in higher energies due to the quality of reconstructed images.  This may considerably influence our estimation of the evaporated mass. To assess the magnitude of this influence, we also performed additional estimations. In the first one, the column density - altitude relations were smoothed using a moving average filter (a span of three to five points) before the subtraction. In the second, we fitted exponential functions to the column density - altitude relations and then the fitted curves were subtracted. This approach seems to be the best way to overcome the noise problem and estimate the evaporated mass more accurately; however, we need to stress that for high column densities (high energies), the fitted curves were constrained with only a few noisy data points. The subtraction of curves gave similar results as the one obtained with the use of smoothed relations. Values of the maximum column mass density change estimated with the use of different methods allowed us to estimate the range of the evaporated mass amount. The range was from $2.7\times10^{13}\,{\rm g}$ to $4.0\times10^{14}\,{\rm g}$.

We compared this range with three other estimations. In each estimation of mass, we took the following plasma abundances: H - 0.92 and He - 0.08 (number fraction of the elements). {Firstly, we estimated the evaporated mass from HD simulations. Depending on the time range used, we obtained values from $2.3\times10^{13}\,{\rm g}$ to $3.3\times10^{13}\,{\rm g}$. Secondly, we estimated the mass of the flare loop-top source using two methods under the assumption that the majority of the loop-top material was transported to the corona from the chromosphere via chromospheric evaporation.}

In the first method, we used the size and emission measure of the loop-top source taken from RHESSI images and imaging spectroscopy. Both input parameters were determined for the time interval just after the analysed HXR burst. The source's size was defined by a 50\% intensity isoline relative to the brightest pixel in the RHESSI image. We measured the area, A, of the source projected on the plane of the image using images from the energy range up to 14~keV. The area differs from image to image, but it is contained in the range $60-70~\rm{arcsec}^2$. This spread was considered as uncertainty regarding the size of the source. Typically, loop-top sources have a circular or elliptical shape. If we assume an ellipsoidal shape of the source, then its volume, $V$, can be calculated from a plane-of-image area, A, as follows:

\begin{equation}
  V=\frac{4}{3}\frac{A^{3/2}}{\sqrt{\pi}}
  \label{eq:vol-area}
.\end{equation}

The emission measure, $EM$, of the loop-top source was obtained from a spectral fit. {The fitting was done for a time interval just after the end of the analysed main HXR burst - 19:04:48 to 19:04:56~UT.} As a result, we got $EM = (4.53 \pm 0.96)\times10^{48} \rm{cm}^{-3}$. Next, having the source volume and emission measure, we calculated the electron number density. Then, assuming fully ionised plasma and the abundance of elements as given above, the mass of the source was determined to be {in the range $(4.45-5.67)\times10^{13}\,{\rm g}$}. The uncertainty of the mass is due to the uncertainty of the estimated size of the loop-top source and its emission measure. {This range can be higher than the actual mass delivered to the loop-top source via evaporation induced by the analysed HXR burst. The difference would come from the fact that there was some mass in the loop-top source before the burst. We calculated this pre-burst mass in the same way as above and the obtained result is $(0.12-0.82)\times10^{13}\,{\rm g}$. Thus, the estimated mass of the loop-top source, after subtraction of its pre-burst mass, is in the range $(3.89-5.29)\times10^{13}\,{\rm g}$, which is consistent with the estimations of the evaporated mass, that is to say is in the range of the mass calculated from the energy-altitude relation and it is close to the mass estimated from HD simulations.} 

The second method is a set of scaling laws describing how the size and physical properties of a loop-top source vary with its altitude above the photosphere \citep{pres2007}. {The altitude was determined from observations. The size of the source can be computed from the following scaling law:}

\begin{equation}
  \log(A)=1.13\log(h)+7.68,
  \label{eq:sc-law1}
\end{equation}

{where $A$ is the area of the loop-top source on the image plane (in square centimetres) and $h$ is the altitude of the source (in centimetres). For the analysed source, we obtained $A\approx 50~\rm{arcsec}^2$, which is in quite good agreement with the area of the source measured in RHESSI images. The volume of the source, V, was calculated from the area, A, in the same way as above (Equation~\ref{eq:vol-area}). The second scaling law that we used allowed us to calculate electron number density, $N_e$. This scaling law depends on a flare magnitude and for X-class flares it looks as follows:}

\begin{equation}
  \log(N_e)=-0.65\log(h)+17.49
  \label{eq:sc-law2}
.\end{equation}

{The set of Equations~\ref{eq:vol-area}-\ref{eq:sc-law2} allowed us to calculate the total number of electrons in the loop-top source. Next, the mass of plasma contained in the source was computed, assuming fully ionised plasma and the same abundance of elements as in the first method (He: 0.92, He: 0.08), and the result is $2\times10^{14}\,{\rm g}$.}

We should note that the scaling laws apply at the specific moment in a flare evolution - the maximum in the GOES 1-8 {\AA} band. The maximum of the analysed flare occurred around 19:08~UT, after three additional HXR bursts which might further supply the loop-top source with mass. Thus, the mass derived from the scaling laws should be considered as the upper estimate of mass delivered to the source {during} the analysed HXR burst.

\section{Conclusions}

In this paper, we presented the analysis of plasma dynamics in a small loop observed in the  SOL2002-08-03T19:07 event. The entire geometry of the event is more complex than for a simple, single-loop. We observed a large arcade of loops simultaneously. However, the RHESSI imaging allowed us to separate HXR emissions arising from the small loop. Thus, we were able to compare observations with the numerical hydrodynamical 1D model of a single loop. 

The dynamics was investigated with the use of energy-altitude relations derived from RHESSI images. Treating non-thermal electrons as a tool, we obtained useful information about plasma distribution and how it changes inside the flaring loop. Extracting spectral parameters of observed sources, we were able to investigate the evolution of purely non-thermal sources and compared them to HD modelling results. We found that the energy-altitude of the evolution of non-thermal sources agree with column density changes within the flaring loop. These changes seem to be dominated by density changes, while a power-law index of an electron beam spectrum is less important. This result is in opposition to the flare analysis done by \cite{oflannagain2013}. We cannot distinguish if such a difference is a real difference between the nature of the two flares until a larger group of events is analysed. 

Purely non-thermal sources visible in SOL2002-08-03T19:07 changed their positions with velocities up to 150~km/s. This value is a typical one, comparable to other observations of chromospheric evaporation, for example \citep{antonucci1982, antonucci1983, landi2003, milligan2006, liu2006, nitta2012, li2017}. However, with our methodology, we measured the evaporation front, while spectroscopy-based methods give information about the whole velocity field in the spectrometer slit's FOV. Moreover, we could obtain a direct estimation of the evaporated mass amount as we measured the position and size of the emitting source from the same image. The obtained mass $7.3\times10^{13}\,{\rm g}$ (a possible range of values is from $2.7\times10^{13}\,{\rm g}$ to $4.0\times10^{14}\,{\rm g}$) is comparable to the mass derived from the HD modelling: $2.3\times10^{13}\,{\rm g}$ to $3.3\times10^{13}\,{\rm g}$. Moreover, we estimated additional mass that occurred in the loop-top by an emission measure time evolution analysis, {and we got $(3.89-5.29)\times10^{13}\,{\rm g}$.} Such consistency of the results supports the usability of the presented methodology.

 {In the first paper considering chromospheric evaporation from RHESSI data, \cite{liu2006} asked questions about the nature of the moving X-ray sources. Firstly, whether they could be characterised as thermal emission from the evaporated hot plasma or as non-thermal emission from the precipitating electrons, or a mixture of both. Secondly, whether they could be related to MHD waves or evaporation fronts.} The analysis of the SOL2002-08-03T19:07 flare allowed us to partially answer these questions. Namely, we see that high-energy, non-thermal sources (above 20~keV in this case) fit the column density changes obtained from the hydrodynamical model perfectly. In lower energies we see a mixture changing the location (upward motion), but the relation to column density change is not straightforward.

The non-thermal foot point sources, if observed with a time resolution of the order of a few seconds, may give valuable information about abrupt density changes within the chromosphere. This opens the possibility for a detailed analysis of mass motions within chromospheric regions with the use of instruments expected in the near future, such as  STIX \citep{krucker2020} onboard Solar Orbiter \citep{muller2020} and HXI \citep{zhang2019} onboard the ASO-S mission \citep{gan2019}. For example, downward moving chromospheric condensation \citep{fisher1985}, occurring simultaneously with chromospheric evaporation (up-flow), may produce a low-density region within the chromosphere. Such a region, existing several {seconds}, may be seen in an energy-altitude relation of HXR sources if they are reconstructed with high time and energy resolutions.

\section*{Acknowledgements}
We thank the anonymous Referee for many valuable comments
and remarks, which have improved this article. We also thank the RHESSI Team for their hard work. This work was supported by the National Science Centre, Poland grants, number 2015/19/B/ST9/02826 and 2020/39/B/ST9/01591.

\bibliographystyle{aa}

\end{document}